\documentclass[12pt]{article}

\usepackage{a4wide, setspace,amsmath,amssymb,graphics,placeins}
 
\title{Self-avoiding walks and polygons on \\
quasiperiodic tilings} 
 
\author{A.~N.~Rogers$^{1)}$, C.~Richard$^{2)}$ and A.~J.~Guttmann$^{1)}$\\
\\
${}^{1)}$Department of Mathematics and Statistics\\
University of Melbourne, Victoria 3010, Australia\\
\\
${}^{2)}$Institut f\"ur Mathematik, Universit\"at Greifswald\\
Jahnstr. 15a, 17487 Greifswald, Germany}

\begin{document}

\maketitle

\begin{abstract}
We enumerate self-avoiding walks and polygons, counted by perimeter,
on the quasiperiodic rhombic Penrose and Ammann-Beenker tilings,
thereby considerably extending previous results.  In contrast to
similar problems on regular lattices, these numbers depend on the
chosen initial vertex.  We compare different ways of counting and
demonstrate that suitable averaging improves convergence to the
asymptotic regime.  This leads to improved estimates for critical
points and exponents, which support the conjecture that self-avoiding
walks on quasiperiodic tilings belong to the same universality class
as self-avoiding walks on the square lattice.  For polygons, the
obtained enumeration data does not allow us to draw decisive
conclusions about the exponent.
\end{abstract}

\section{Introduction}

Quasiperiodic tilings are most widely used for the description of
quasicrystals.  With appropriate atomic decorations of the vertices,
they serve as structure models which explain physical properties of
quasicrystals \cite{GKTU00}.  From a theoretical point of view, they
are idealisations of real substances on which the usual models of
statistical physics like the Ising model may be studied \cite{RGS99,
R02, L93}.  Quasiperiodic tilings arose before the discovery of
quasicrystals, however, more as an object of aesthetic interest in
geometry \cite{P74, KN84}.

From a combinatorial point of view, they provide an interesting
example of a non-periodic yet structured graph where typical problems
of combinatorics like the counting of objects on a lattice become more
complex.  This is fundamentally different from counting problems on
semiregular lattices, where the underlying translational invariance is
still present \cite{JG98}, and also from self-similar graphs, where
the self-similarity allows for the solvability of some counting problems.

Consider for example the counting problem of $n$-step walks on a
quasiperiodic tiling.  This number depends on the chosen starting
point.  On a lattice, this phenomenon does not occur due to
translation invariance.  Questions arise, such as if the universal
properties of the walks like critical exponents \cite{MS93} are
changed for quasiperiodic tilings\footnote
{
For ferromagnetic spin systems on quasiperiodic graphs, a heuristic
criterion determines whether its behaviour is different from the
lattice situation \cite{L93}.
}
, and how different ways of counting affect asymptotic properties.
The first question has been investigated for self-avoiding walks
(SAWs) and self-avoiding polygons (SAPs).  These are walks or loops
which do not visit the same vertex twice.  Extrapolation of exact
enumeration data for a number of quasiperiodic tilings \cite{B93,
RGS99} indicates that the critical exponents $\gamma$ for SAWs and
$\alpha$ for SAPs are consistent with the corresponding values on
regular lattices $\alpha=1/2$ and $\gamma=43/32$.  In \cite{LI92}, the
related problem of SAWs on Penrose random tilings \cite{Hen99} has
been studied by Monte Carlo simulations, the results suggesting the
same mean square displacement exponent as in the lattice situation.
The studies \cite{B93, RGS99} suffered however from strong finite-size
effects due to the relatively short series data available.  This is
mainly due to the fact that the finite-lattice method \cite{E96},
being the most successful method known to date for walk enumeration on
regular lattices \cite{CGE93, CG96}, cannot easily be applied here,
and so we used the slower method of direct counting.  The problem in
applying the FLM is discussed in Section 3 below, see also \cite{R02}.
Another reason for the pronounced finite-size behaviour is that the
number of walks or loops depends on the chosen starting point.  One
might suspect that suitable averaging over different starting points
reduces these effects, leading to behaviour comparable to the square
lattice case.  In this paper, we analyse three different methods of
counting in detail.  Whereas the first one depends on a chosen vertex
of the tiling, the last two are averages over the whole tiling.  
\begin{itemize}
\item 
{\it Fixed origin walks.} We count the number of $n$-step
self-avoiding walks emanating from a given vertex.  This number
depends on the chosen vertex.
\item
{\it Mean number of walks.}
We count all translationally inequivalent $n$-step self-avoiding walks
which may occur anywhere in the infinite tiling, weighted by their
occurrence probability.  For tilings with quasicrystallographic
$k$-fold symmetries, these probabilities are numbers in the underlying
module $\mathbb Z[e^{2 \pi i /k}]$.  This leads to a generating
function which has non-integer coefficients.
\item
{\it Total number of walks.}
Here we count the number of translationally inequivalent $n$-step
self-avoiding walks which may occur anywhere in the tiling.  This
number is bigger than the number of fixed origin walks, and by
definition, takes into account vertices over the whole tiling.
\end{itemize}
Note that two self-avoiding walks (polygons) are translationally
equivalent iff they have, up to a translation, the same vertex
coordinates.  For self-avoiding polygons, we will employ the second
and the third method of counting.  We do not distinguish between
different fillings of the interior of the polygon.  For SAPs, the
second method has been implemented previously \cite{RGS99} to obtain
the high temperature expansion of the Ising model.  The mean number of
SAPs up to length $2n=18$ has been determined on the Ammann-Beenker
tiling \cite{AGS92, K95} and on the rhombic Penrose tiling \cite{P74,B81}.

We counted SAWs and SAPs on the Ammann-Beenker tiling and the rhombic
Penrose tiling and compared different counting schemes, thereby
extending and generalising the previous approaches to counting SAWs
\cite{B93} and SAPs \cite{RGS99}.  Generally speaking, averaging
reduces oscillation of data due to finite size effects, providing
improved estimates for critical points and critical exponents.  Within
numerical accuracy, we cannot rule out the universality hypothesis
that SAWs on the Ammann-Beenker and on the rhombic Penrose tilings
have the same exponents as on the square lattice.  The data for the
total number of walks (polygons) gives a different exponent,
reflecting the fact that the number of patches grows quadratically
with the patch size, in contrast to the lattice case
\cite{LP01,Le02}.  The limited quantity of SAP data does not allow us
to draw decisive conclusions about exponents.

This paper is organised as follows.  The next chapters describe the
algorithms used for the generation of the tilings and for the
computation of the numbers of walks and their mean values.  The
following chapter is devoted to the asymptotic analysis of the series
and to a comparison of the different approaches.  This is concluded by
a discussion of possible future work.

\section{Graph generation}

Quasiperiodic tilings in $\mathbb R^d$ may be obtained by projecting
certain subsets of lattices from a higher-dimensional space $\mathbb
R^n$ into $\mathbb R^d$.  This is described by a {\it cut-and-project
scheme}, summarised in the following diagram.

\begin{displaymath}
\renewcommand{\arraystretch}{1.5}
\begin{array}{ccccc}
& \pi_{||} & & \pi_{\perp} & \vspace*{-1.5ex} \\
 E_{||}\simeq\mathbb R^{d} & \longleftarrow & E=\mathbb R^{n} & \longrightarrow 
& E_{\perp}\simeq\mathbb R^m \\ 
\cup  & \mbox{\raisebox{-1.5ex}{\footnotesize
     \textnormal{1--1}}}\!\!\!\!\nwarrow\;\; & \cup &  
\;\;\nearrow\!\!\!\!\mbox{\raisebox{-1.5ex}{\footnotesize \textnormal{dense}}}
& \cup\\
L_{||}=\pi_{||}(L) & & L \mbox{ lattice} & & W \mbox{ polytope}
\end{array}
\end{displaymath} 
It consists of a Euclidean vector space $E$, together with orthogonal projections 
$\pi_{||}$ and $\pi_{\perp}$.
The vector spaces $E_{||}=\pi_{||}(E)$ and $E_{\perp}=\pi_{\perp}(E)$ are called
{\it direct} and {\it internal} space, respectively.
Let $L\subset E$ be a lattice.
The projections are such that $\pi_{||}|_{L}$ is one-to-one and 
$\pi_{\perp}(L)$ is dense in $E_{\perp}$ (or dense in some subspace of $E_{\perp}$).
Let $W\subset E^{\perp}$ be a polytope 
(or a finite union of polytopes).
The set $W$ is also called the  {\it acceptance window}.
The set of tiling vertices $\Lambda(W)$ is defined by
\begin{equation}
\Lambda(W) = 
\{ \mathbf{x}_{||} \in L_{||} \, | \, \mathbf{x}\in L \mbox{ and } 
\mathbf{x}_{\perp} \in W\}.
\end{equation}
The edges of the tiling are defined by the following rule:
The tiling vertices $\pi_{||}(\mathbf{x})$ and 
$\pi_{||}(\mathbf{y})$ are adjacent iff the lattice vectors 
$\mathbf{x}$ and $\mathbf{y}$ are adjacent.

\begin{figure}[ht]
\center{\resizebox{.5\textwidth}{!}{\includegraphics{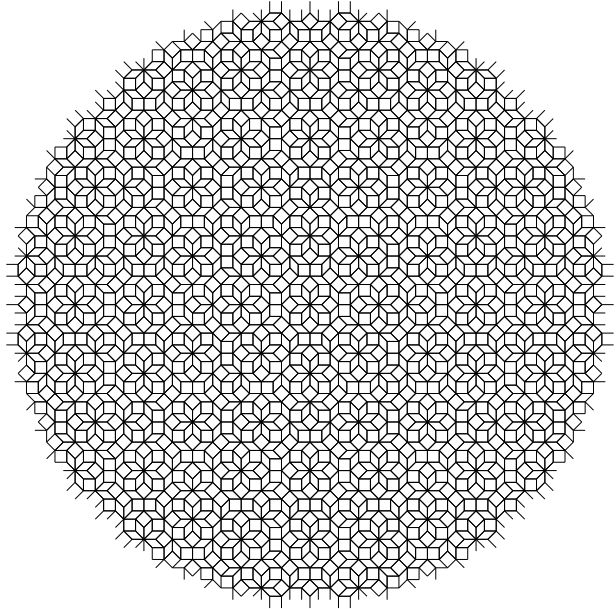}}}
\caption{A patch of the Ammann-Beenker tiling.}\label{Fi:ABQL}
\end{figure}

For the Ammann-Beenker tiling \cite{AGS92, K95}, we have $n=4$ and $d=m=2$.
The lattice is $L=\mathbb Z^4$.
The projections $\pi_{||}$ and $\pi_{\perp}$ are defined as follows.
For $\mathbf{x}\in\mathbb R^n$, we set
\begin{eqnarray}
\mathbf{x}_\parallel &=& \left( \begin{matrix}
  1 & \cos \frac{\pi}{4} & \cos \frac{2\pi}{4} & \cos \frac{3\pi}{4} \\
  0 & \sin \frac{\pi}{4} & \sin \frac{2\pi}{4} & \sin \frac{3\pi}{4}
  \end{matrix} \right) \mathbf{x}, \\
\mathbf{x}_\perp &=& \left( \begin{matrix}
  1 & \cos \frac{3\pi}{4} & \cos \frac{6\pi}{4} & \cos \frac{9\pi}{4} \\
  0 & \sin \frac{3\pi}{4} & \sin \frac{6\pi}{4} & \sin \frac{9\pi}{4}
  \end{matrix} \right) \mathbf{x}. \nonumber
\end{eqnarray}
The acceptance window $W\subset\mathbb R^m$ is a regular octagon with unit
side length centred at the origin, having edges perpendicular to the axes.
A typical patch is shown in Figure \ref{Fi:ABQL}.

For the rhombic Penrose tiling \cite{P74, B81}, we have $n=5$, $d=2$ and $m=4$.
The lattice is $L=\mathbb Z^5$.
The projections $\pi_{||}$ and $\pi_{\perp}$ are, for
$\mathbf{x}\in\mathbb R^n$, defined by
\begin{eqnarray}
\mathbf{x}_\parallel &=& \left( \begin{matrix}
  1 & \cos \frac{2\pi}{5} & \cos \frac{4\pi}{5} & \cos \frac{6\pi}{5} & \cos \frac{8\pi}{5} \\
  0 & \sin \frac{2\pi}{5} & \sin \frac{4\pi}{5} & \sin \frac{6\pi}{5} & \sin \frac{8\pi}{5}
  \end{matrix} \right) \mathbf{x}, \\
\mathbf{x}_\perp &=& \left( \begin{matrix}
  1 & \cos \frac{4\pi}{5} & \cos \frac{8\pi}{5} & \cos \frac{12\pi}{5} & \cos \frac{16\pi}{5} \\
  0 & \sin \frac{4\pi}{5} & \sin \frac{8\pi}{5} & \sin \frac{12\pi}{5} & \sin \frac{16\pi}{5} \\
  1 & 1 & 1 & 1 & 1
  \end{matrix} \right) \mathbf{x}. \nonumber
\end{eqnarray}

\begin{figure}[ht]
\center{\resizebox{.5\textwidth}{!}{\includegraphics{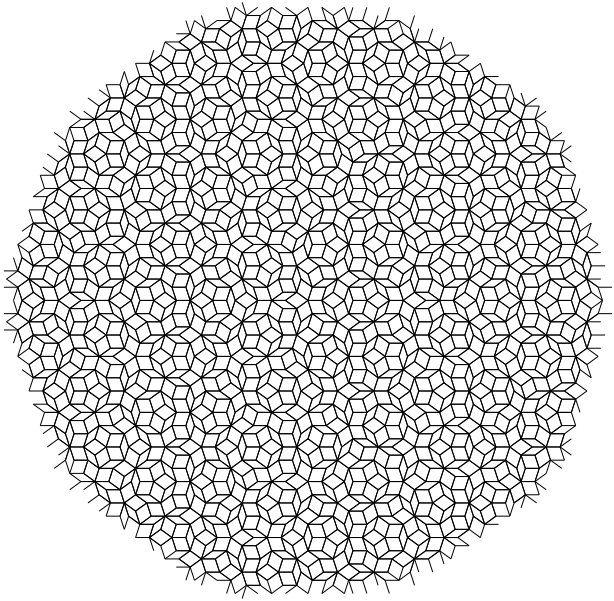}}}
\caption{A patch of the rhombic Penrose tiling.}\label{Fi:PRQL}
\end{figure}

The acceptance window $W\subset\mathbb R^m$ is made up of four regular
pentagons in the planes $\mathbf{x}_{\perp3} = 0,1,2,3$.  The pentagons in the
$0$ and $3$ $x_3$-planes have unit side length and the others
have side length
$2\cos \frac{\pi}{5}$.  Each pentagon is centred at $\mathbf{x}_{\perp1} = 0$,
$\mathbf{x}_{\perp2} = 0$.  Pentagons $0$ and $2$ have an edge crossing the
positive $\mathbf{x}_{\perp1}$ axis at right angles while pentagons $1$ and $3$
are rotated through $\frac{\pi}{5}$.
A typical patch is shown in Figure \ref{Fi:PRQL}.

We remark in passing that a more natural embedding of the rhombic Penrose tiling 
is the root lattice $A_4$, see also \cite{BJKS90}, but $\mathbb{Z}^5$ is more 
convenient for computations.
Moreover, the Ammann-Beenker tiling and the rhombic Penrose tiling may be 
alternatively defined by inflation rules for their prototiles \cite{BJ90,GS87}.

\FloatBarrier
\section{Enumeration}

A self-avoiding walk on a graph is a path, beginning at an origin
vertex, which never visits a vertex more than once. 
The SAW on the lattice $\mathbb Z^d$ is a well-studied object (see, for
example, \cite{MS93,LSW02}). 
The number $c_n$ of translationally inequivalent $n$-step walks on a
regular lattice is clearly independent of the choice of origin vertex.
Hence this series is representative of the entire lattice.
The enumeration of SAWs on non-periodic tilings introduces complications to
the interpretation of the series $C(x) = \sum_{n \ge 0} c_n x^n$. This
is because the possible origin vertices produce an infinite range of
different series $C(x)$.
Each origin produces a different series which is representative only
of that vertex's immediate neighbourhood in the tiling.
The question then is, how do we obtain a SAW series which is representative of
the whole tiling?  In this paper we adopt three different approaches to
enumerating SAWs on quasiperiodic tilings, as described in the introduction.
They are:
\begin{itemize}
  \item \emph{Fixed origin walks.} 
  \item \emph{Mean number of walks.} 
  \item \emph{Total number of walks.} 
\end{itemize}

\FloatBarrier
\subsection{Fixed origin walks}

We take a random selection of origin vertices $\mathbf{x} \in L$ (if
$\mathbf{x}_\perp \notin W$ the vertex is not a suitable choice and is ignored).
For each suitable origin, we generate the neighbourhood of the vertex, including
all vertices up to some Euclidean distance $N$ away.  Two such neighbourhoods are
shown in Figure \ref{Fi:ABQL} and Figure \ref{Fi:PRQL}.
We enumerate all SAWs from the origin up to length $n$ in the
neighbourhood using backtracking \cite{S92}.
This takes time proportional to the number of walks $c_n$.
Unfortunately, the transfer matrix approaches used to enumerate SAWs on regular
two-dimensional lattices in less time cannot be used on this problem without major
adaptation:
Since a SAW of $N$ steps may reach a vertex $N$ steps from the origin,
we would need to consider every possible tiling patch of radius $N$.
The finite-lattice method's transfer matrix stage would then need to be
adapted to each tiling patch, or generalised to handle them all.

\begin{figure}[ht]
\begin{tabular}{c c c c}
\resizebox{.2\textwidth}{!}{\includegraphics{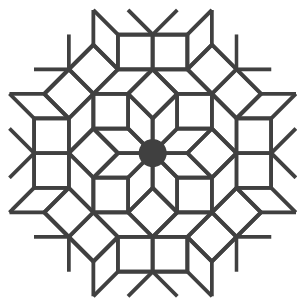}} &
\resizebox{.2\textwidth}{!}{\includegraphics{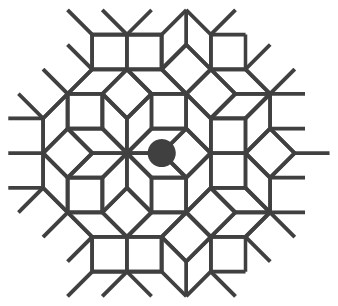}} &
\resizebox{.2\textwidth}{!}{\includegraphics{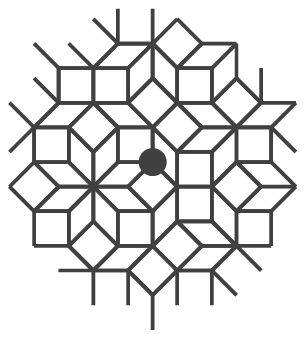}} &
\resizebox{.2\textwidth}{!}{\includegraphics{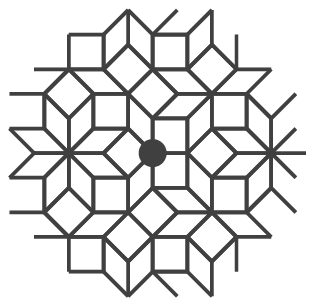}} \\
$\mathbf{x}_\perp=(0,0)$ & $\mathbf{x}_\perp=(1,0)$ & $\mathbf{x}_\perp=(1-\frac{\sqrt{2}}{2},\frac{\sqrt{2}}{2})$ & $\mathbf{x}_\perp=(1-\sqrt{2},0)$ \\
\end{tabular}
\caption{The actual Ammann-Beenker neighbourhoods chosen for the enumerations.}\label{Fi:ABpatchs}
\end{figure}

\begin{table}[ht]
\center{\scalebox{1.0}{\parbox{\textwidth}{
\begin{tabular}{|r|r|r|r|r|} \hline
$n$ & $\mathbf{x}_\perp=(0,0)$ & $\mathbf{x}_\perp=(1,0)$ & $\mathbf{x}_\perp=(1-\frac{\sqrt{2}}{2},\frac{\sqrt{2}}{2})$ & $\mathbf{x}_\perp=(1-\sqrt{2},0)$ \\ \hline
0 & 1 & 1 & 1 & 1 \\
1 & 8 & 3 & 4 & 5 \\
2 & 16 & 13 & 12 & 16 \\
3 & 48 & 34 & 46 & 42 \\
4 & 144 & 108 & 108 & 152 \\
5 & 448 & 292 & 374 & 388 \\
6 & 1088 & 952 & 976 & 1194 \\
7 & 3680 & 2458 & 3042 & 3412 \\
8 & 9584 & 7746 & 8330 & 9678 \\
9 & 28336 & 21348 & 24556 & 27218 \\
10 & 82960 & 61478 & 68376 & 79150 \\
11 & 225408 & 177230 & 197820 & 217562 \\
12 & 657536 & 495808 & 554108 & 628996 \\
13 & 1834768 & 1412152 & 1576464 & 1741464 \\
14 & 5140752 & 3985706 & 4400920 & 4968606 \\
15 & 14584112 & 11125408 & 12531794 & 13724682 \\
16 & 40222672 & 31617786 & 34541864 & 39209054 \\
17 & 114683280 & 87149372 & 98846548 & 107503768 \\
18 & 313146848 & 248799302 & 270221012 & 306845714 \\
19 & 896810944 & 680172768 & 773046904 & 840463852 \\
20 & 2437468000 & 1943692238 & 2109562128 & 2386875508 \\
21 & 6958267152 & 5303535884 & 6011045200 & 6548653714 \\
22 & 18981078176 & 15086983820 & 16431248782 & 18500898140 \\
23 & 53728620912 & 41295324398 & 46538635588 & 50883461478 \\
24 & 147472084608 & 116624466842 & 127704810544 & 142927122532 \\
25 & 413887940176 & & & \\ \hline
\end{tabular}
}}
\caption{The number of $n$-step fixed origin SAWs for various 
starting points $\mathbf{x}_\perp$ in the Ammann-Beenker tiling, with starting point coordinates 
$(x,y)$ given in the internal space.}\label{Ta:VOAB}
}
\end{table}

\begin{figure}[ht]
\begin{tabular}{c c c c}
\resizebox{.2\textwidth}{!}{\includegraphics{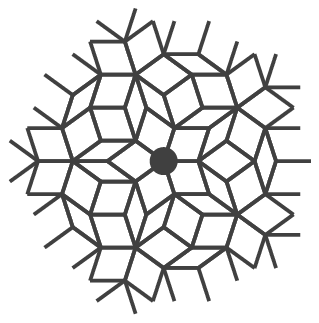}} &
\resizebox{.2\textwidth}{!}{\includegraphics{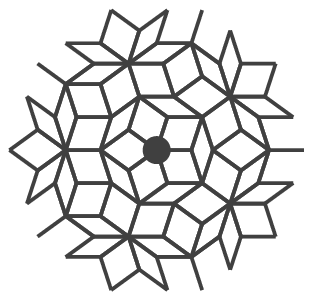}} &
\resizebox{.2\textwidth}{!}{\includegraphics{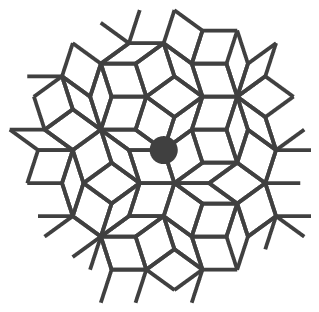}} &
\resizebox{.2\textwidth}{!}{\includegraphics{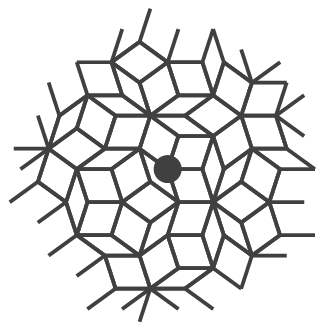}} \\
$\mathbf{x}_\perp=(0,0,1)$ & $\mathbf{x}_\perp=(0,0,0)$& 
$\mathbf{x}_\perp=(0.5,0.5,1)$ & $\mathbf{x}_\perp=(0.25,0.5,0)$\\
\end{tabular}
\caption{The actual rhombic Penrose neighbourhoods chosen for the enumerations.}\label{Fi:PRpatchs}
\end{figure}

\begin{table}[ht]
\center{\scalebox{1.00}{\parbox{\textwidth}{
\begin{tabular}{|r|r|r|r|r|} \hline
$n$ & $\mathbf{x}_\perp=(0,0,1)$ & $\mathbf{x}_\perp=(0,0,0)$& 
$\mathbf{x}_\perp=(0.5,0.5,1)$ & $\mathbf{x}_\perp=(0.25,0.5,0)$\\ \hline
0 & 1 & 1 & 1 & 1 \\
1 & 5 & 5 & 5 & 4 \\
2 & 20 & 10 & 14 & 12 \\
3 & 40 & 40 & 50 & 46 \\
4 & 160 & 130 & 130 & 112 \\
5 & 450 & 310 & 406 & 394 \\
6 & 1170 & 1140 & 1177 & 938 \\
7 & 4000 & 2680 & 3316 & 3416 \\
8 & 9480 & 9360 & 9723 & 7866 \\
9 & 32910 & 23150 & 27356 & 27312 \\
10 & 76090 & 72520 & 77747 & 66150 \\
11 & 262250 & 196980 & 224102 & 215924 \\
12 & 619460 & 555290 & 615549 & 545062 \\
13 & 2050500 & 1646990 & 1812802 & 1698548 \\
14 & 5052310 & 4292010 & 4869786 & 4411293 \\
15 & 15828550 & 13403280 & 14455725 & 13367278 \\
16 & 41103090 & 33637420 & 38524509 & 35243859 \\
17 & 121759470 & 106779600 & 114089288 & 105117832 \\
18 & 331072990 & 265198150 & 304434061 & 279216083 \\
19 & 937563530 & 840669610 & 894584372 & 825140032 \\
20 & 2642381430 & 2092703550 & 2399386239 & 2199738033 \\
21 & 7227151280 & 6573888100 & 6988332717 & 6459329037 \\
22 & 20931973090 & 16491425740 & 18844561759 & 17267339059 \\
23 & 55793302330 & 51185968460 & 54473434666 & 50419312152 \\
24 & 164764171030 & 129673789110 & 147471723662 & 135162732506 \\ \hline
\end{tabular}
}}}
\caption{The number of fixed origin $n$-step SAWs for various starting points 
$\mathbf{x}_\perp$ in the rhombic Penrose tiling, with starting point coordinates $(x,y,z)$ given
in the internal space.}\label{Ta:VOPR}
\end{table}

If we apply this method to counting walks on a regular lattice it is clear that
we would always produce the usual SAW series for that lattice.  If each of the
series in Table \ref{Ta:VOAB} and Table \ref{Ta:VOPR} showed lattice consistent
properties, it would be a good indication that these properties belong to the entire
tiling. The actual neighbourhoods chosen for the enumerations are shown in Figure \ref{Fi:ABpatchs} and Figure \ref{Fi:PRpatchs} for the Ammann-Beenker and Penrose tilings respectively.

\FloatBarrier
\subsection{Mean number of walks}

Given that a pair of vertices from $\Lambda(W)$ are adjacent if and only if they are
adjacent in $L$, we see that the neighbours of a vertex with image
$\mathbf{x}_\perp$ can be found by sequentially adding the $E_\perp$ image of
all possible edges in $L$ to $\mathbf{x}_\perp$ and testing if the new points
lie in $W$.  If it does the adjacent vertex exists in $\Lambda(W)$.  
By recursively checking all possible neighbours of a vertex, all possible walks 
on the lattices will be found.

Given an origin vertex $x^0$ in $\Lambda(W)$, we know its $E_\perp$ image 
$x^0_\perp$ must lie somewhere in $W$, i.e. $x^0_\perp \in W^0 = W$ ($W^n$ is the region
$x^0_\perp$ can lie in given our knowledge of the $n$ steps in the walk).  If
we take a step $s$ (with projection $s_\perp$ onto $E_\perp$) to a possible
adjacent vertex $x^1$, then we know $x^1_\perp = x^0_\perp + s_\perp$.
Furthermore if $x^1 \in \Lambda(W)$ is true, $x^1_\perp \in W$.  Hence
$W^1 = (W \cap (W^0 + s_\perp)) - s_\perp$.  The probability that the
step $s$ is possible from a random $x^0$ is given by the ratio between the areas of
$W^1$ and $W$.

\begin{figure}
\center{\resizebox{.7\textwidth}{!}{\rotatebox{270}{\includegraphics{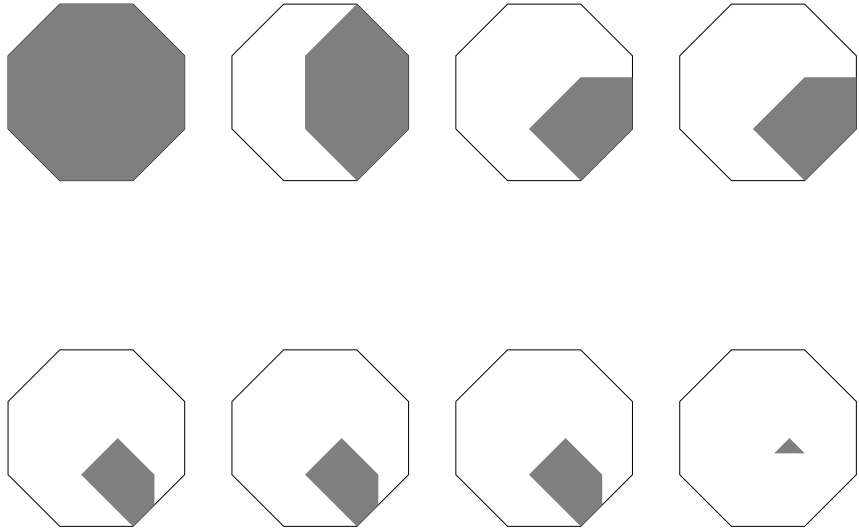}}}}
\caption{Examples of $W^0$ \ldots $W^7$ for a particular walk on the 
Ammann-Beenker tiling.}\label{Fi:ABW8}
\end{figure}

Extending this to a walk of length $n$ with steps $s^i, i = 1 \dots n $,
$W^k = (W \cap (W^{k-1} + \sum_{i=1}^ks^i_\perp)) - \sum_{i=1}^ks^i_\perp,$ 
the probability of the walk existing is the ratio of the areas of $W^n$ and $W$.
For example, consider the shaded $W^i$ in Figure \ref{Fi:ABW8}, for the particular
walk in the Ammann-Beenker tiling which steps west, south, south-west,
north-west, west, north then north.
For the Ammann-Beenker tiling, these probabilities are of the form $a+b\lambda$, 
where $\lambda=1+\sqrt{2}$ and $a,b\in\mathbb Q$.

Adding the self-avoiding constraint and summing the probabilities results in the
expected number of SAWs beginning at a random origin.  Again note that applying
this method to a regular lattice would result in the usual SAW series.  Due to the
extra complexity of the rhombic Penrose tiling acceptance window, the area
calculations are more involved, see also \cite{RGS99}.
They lead to mean numbers of the form $a+b\tau$, where $\tau=(1+\sqrt{5})/2$ 
is the golden number, and $a,b\in\mathbb Z$.
The rhombic Penrose tiling also allows steps in ten directions, 
more than the Ammann-Beenker tiling's eight.  
These facts combine to allow greater length series to be computed on the Ammann-Beenker tiling.

\begin{table}[ht]
\center{\scalebox{1.0}{\parbox{\textwidth}{
\begin{tabular}{|r|r|r|} \hline
$n$ & Ammann-Beenker & rhombic Penrose\\ \hline
0 & 1 
& 1 \\
1 & 4 
& 4 \\
2 & 52-16$\lambda$   
& 62 -30$\tau$ \\
3 &  80-16$\lambda$ 
& -4 + 28$\tau$ \\
4 &  444-134$\lambda$ 
& 914 -488$\tau$ \\
5 &  1280-380$\lambda$ 
& -820 + 732$\tau$ \\
6 &  4492-1430$\lambda$ 
& 13842 -7894$\tau$ \\
7 &  10848-3248$\lambda$ 
& -17732 + 12860$\tau$ \\
8 &  60988-21700$\lambda$ 
& 173876 -101988$\tau$ \\
9 &  89800-27036$\lambda$ 
& -255784 + 173720$\tau$ \\
10 &  643248-237732$\lambda$ 
& 1923078 -1143988$\tau$ \\
11 &  979776-324200$\lambda$ 
& -3149856 + 2073192$\tau$ \\
12 &  5486960-2043420$\lambda$ 
& 19566548 -11734706$\tau$ \\
13 &  10785736-3819788$\lambda$ 
& -34951044 + 22612992$\tau$ \\
14 &  45253532-16927618$\lambda$ 
& 192557132 -116151274$\tau$ \\
15 &  110294592-40576780$\lambda$ 
& -366912524 + 234803904$\tau$ \\
16 &  375796808-141368464$\lambda$ 
& \\
17 &  1058437232-398339560$\lambda$ 
& \\
18 &  3259350860-1238175678$\lambda$ 
& \\
19 & 9526156024-3632872284$\lambda$ 
& \\
20 &  29127575440-11192322668$\lambda$ 
& \\ 
21 & 81536068712-31337365980$\lambda$ 
& \\
22 & 259724099656-100797073134$\lambda$ 
& \\ \hline
\end{tabular}
}}}
\caption{The mean number of $n$-step SAWs for the Ammann-Beenker tiling and 
the rhombic Penrose tiling, where $\lambda=1+\sqrt2$ and $\tau=(1+\sqrt5)/2$.}
\label{Ta:mean}
\end{table}

\FloatBarrier
\subsection{Total number of walks}

Investigating all possible walks as in the mean number of walks method, we count
instead the number of non-zero contributions to the mean value.  This counts
the number of translationally inequivalent walks with $W^n$ having
positive area or, equivalently, the number of translationally
inequivalent walks which may occur anywhere in the tiling.  Again we note that applying
this method to a regular lattice gives the usual SAW series.

\begin{table}[ht]
\center{\scalebox{1.0}{\parbox{\textwidth}{
\begin{tabular}{|r|r|r|} \hline
$n$ & Ammann-Beenker & Penrose\\ \hline
0 & 1 & 1 \\
1 & 8 & 10 \\
2 & 56 & 90 \\
3 & 288 & 560 \\
4 & 1280 & 2800 \\
5 & 5344 & 12060 \\
6 & 20288 & 48520 \\
7 & 74192 & 182000 \\
8 & 260336 & 658300 \\
9 & 892800 & 2282400 \\
10 & 2976512 & 7749440 \\
11 & 9828256 & 25634920 \\
12 & 31758112 & 83615140 \\
13 & 101847216 & 268113660 \\
14 & 322240144 & 850895040 \\
15 & 1012048208 & 2668534600 \\
16 & 3147031584 & \\
17 & 9732815728 & \\
18 & 29852932384 & \\
19 & 91182029360 & \\
20 & 276695822928 & \\
21 & 836719766336 & \\
22 & 2516664888416 & \\ \hline
\end{tabular}
}}}

\caption{The total number of $n$-step SAWs for the Ammann-Beenker tiling and 
the rhombic Penrose tiling.}\label{Ta:any}
\end{table}

\FloatBarrier
\subsection{Self-avoiding polygons}

A self-avoiding polygon is equivalent to a self-avoiding walk in which the
initial and final vertices are adjacent.
In the enumeration of self-avoiding polygons, we do not distinguish
between polygons having, up to a translation, the same boundary but different fillings of the interior.

SAPs may be enumerated in the same manner as we enumerate SAWs.
The additional property of end point adjacency allows the backtracking
algorithm to be pruned earlier.  When enumerating SAPs up to size $N$,
a walk that visits a vertex after $D$ steps that is further than $N-D$ steps from
the origin may be pruned from the search tree.  Such a walk can never form part
of a SAP of  $\le N$ steps.
This was used to extend the length of the rhombic Penrose SAP series. The
Ammann-Beenker SAP series were calculated at the same time as the SAW series
and hence are of the same length.  The extensive run time requirements
precluded further series extension.

For the computation of occurrence probabilities of self-avoiding
polygons, the loop vertices are taken into account, as described in
Section 3.2 for the mean number of walks.  If the self-avoiding
polygon has $n$ loop vertices $x^i\in\Lambda(W)$, where
$i=1,\ldots,n$, the acceptance domain is $W^n=\bigcap_i
(W-x_\perp^i)$, and the occurrence probability is given by the ratio
between the areas of $W^n$ and $W$, see also \cite{RGS99}.

\begin{table}[ht]
\center{\scalebox{1.0}{\parbox{\textwidth}{
\begin{tabular}{|r|r|r|r|r|} \hline
$n$ & mean number & total number & mean number & total number\\ \hline
2 & 4 
& 8 & 4  &  10 \\
4 &  8 
& 48 & 8 &  80\\
6 &  12$\lambda$ 
& 384 &  108-48$\tau$ &840\\
8 &  800-272$\lambda$ 
& 2960 & 240-64$\tau$ &  6480  \\
10 &  2840-880$\lambda$ 
& 21600 & 6192-3364$\tau$ & 49760\\
12 &  28152-9984$\lambda$ 
&  170256 & 25584-13248$\tau$ &394080 \\
14 &  47712-9884$\lambda$ 
& 1322048 &  179200-95340$\tau$ &  3087140\\
16 &  869600-299392$\lambda$ 
& 10194720 & 162976-5440$\tau$ & 24020160\\
18 &  215712+294408$\lambda$ 
& 79960896 & 2704140-1067580$\tau$& 183529440\\
20 &  14980920-3730840$\lambda$ 
& 618248240 & & \\ 
22 &  152588920-47048100$\lambda$ 
& 4726263168 & & \\ \hline
\end{tabular}
}}}

\caption{The mean number of $n$-step SAPs and the total number of SAPs 
for the Ammann-Beenker tiling (first two columns) and for the rhombic Penrose tiling
(last two columns), where $\lambda=1+\sqrt2$ and $\tau=(1+\sqrt5)/2$.}\label{Ta:sap}
\end{table}

\section{Analysis of series}

The various sequences were analysed using standard methods of asymptotic
analysis of power series expansions as described in \cite{G89}.
For self-avoiding walks and polygons, it is easy to prove that the limit
$\lim_{n\to\infty} (c_n)^{1/n}$ exists by use of concatenation
arguments \cite{MS93}. 
We assume the usual asymptotic growth of the sequence coefficients $c_n,$ viz:
\begin{equation}
c_n = A x_c^{-n} n^{\gamma-1}\left[1 + {\cal O}(n^{-\varepsilon})\right] 
\qquad (n\to\infty, 0<\epsilon\le1).
\end{equation}
On the square lattice, there is overwhelming evidence \cite{CG96} 
of the above asymptotic behaviour with $\gamma=43/32$. There is however
no proof of this assumption.
For interesting new developments see \cite{LSW02}.
The above assumption results in an asymptotic growth of the ratios $r_n$
\begin{equation}
r_n=\frac{c_n}{c_{n-1}}=\frac{1}{x_c}
\left[ 1+\frac{\gamma-1}{n}
+ {\cal O}(n^{-1-\varepsilon}) \right]
\qquad (n\to\infty, 0<\epsilon\le1),
\end{equation}
which may be used to extrapolate numerical estimates of $x_c$ and $\gamma$.
Whereas it has been proved for the square lattice that the limit 
$\lim_{n\to\infty} c_n/c_{n-2}$ exists and coincides with 
$x_c^{-2}$ \cite{K63}, 
a similar statement for the ratios $r_n$ is not known.
For some lattices, counterexamples are known \cite{H60}.

\begin{figure}
\center{\resizebox{.6\textwidth}{!}{\includegraphics{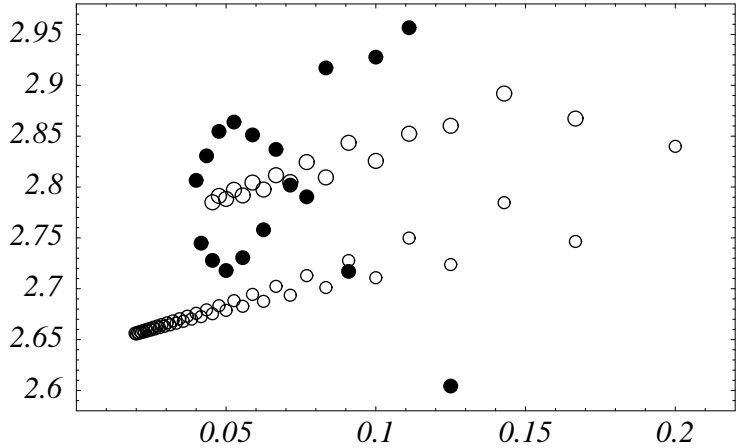}}}
\caption{Ratio plot of $c_n/c_{n-1}$ against $1/n$ for
fixed origin (full circles) and 
mean number (large empty circles) Ammann-Beenker SAWs.
The square lattice SAW data is plotted in small circles for comparison.}
\label{Fi:data}
\end{figure}

Fig.~\ref{Fi:data} shows a plot of the ratios $r_n$ against $1/n$ for 
a typical fixed origin Ammann-Beenker walk (full circles) and for
the ratios of the mean numbers of Ammann-Beenker walks (large empty circles).
We notice that the fixed origin data suffers from dramatic fluctuations,
which are smoothed out by averaging, but are still larger than the corresponding
square lattice data \cite{CG96}, which is shown in small circles.
The oscillating behaviour of the mean number of walks data is due to
an additional singularity of the sequence generating 
function at $x=-x_c$, which, for the case of the square lattice, 
is well understood due to anti-ferromagnetic ordering \cite{CG96}.
To obtain estimates of $x_c$ and $\gamma$, we used the standard method 
described in \cite{G89} and first mapped away the 
singularity on the negative real axis by an Euler transform and then used 
Neville-Aitken series extrapolation.

We also used the method of differential approximants (DAs) \cite{G89}.
The underlying idea is to fit a linear differential equation with polynomial
coefficients to the generating function of the sequence, truncated at some 
order $n_0$.
The critical points and critical exponents of the differential equation
are expected to approximate the critical behaviour of the underlying sequence.
Application of first order and second order DAs has proved useful for the analysis
of square lattice SAWs \cite{G89}.
A first order DA involves fitting the coefficients to the differential
equation $P_1(x)xf'(x) + P_0(x) f(x) = R(x)$ where $P_1(x)$, $P_0(x)$ and 
$R(x)$ are polynomials of degree $N,$ $M,$ and $L$ respectively. We refer
to this as a $[L/M;N]$ DA.
We analysed the approximants $[L/N-1;N]$, $[L/N;N]$, $[L/N+1;N]$ for
$1\le L\le 8$.
We computed estimates for $x_c$ and $\gamma$ by averaging of
the different DA results, given a fixed number of series coefficients $n_0$. 
Since this yields more accurate estimates of $x_c$ and $\gamma$ 
(typically of one digit better) than the ratio method and series extrapolation, 
we list in Table \ref{Ta:diffappAmm} only the results for the DA analysis.
Note that the errors are no strict error bounds, but arise from
averaging over approximants $[L/N_0;N_1]$ for different values of
$N_0,N_1$ and $L$, as explained in \cite{G89}.

\begin{table}[ht]
\center{\scalebox{1.0}{\parbox{\textwidth}{
\begin{tabular}{|c|r|r|r|r|r|} \hline
 & mean no. & $(0,0)$ & $(1-\frac{\sqrt{2}}{2},\frac{\sqrt{2}}{2})$ & $(1,0)$ & $(1-\sqrt{2},0)$ \\ \hline
$x_c$ & 0.36414(18)& 0.3659(14) & 0.3644(13) & 0.3644(21) & 0.3657(16)\\ \hline
$\gamma$ & 1.325(19) & 1.45(16) & 1.30(14) & 1.32(17) & 1.46(17) \\
 \hline
\end{tabular}
}}}
\caption{Estimates of $x_c$ and $\gamma$ for Ammann-Beenker SAWs.
Numbers in brackets denote the uncertainty in the last two digits.}
\label{Ta:diffappAmm}
\end{table}

As suggested by the ratio plot, the estimate using the data for the mean number 
of walks yields the most precise estimates, which are, however, one order
of magnitude in error worse than the corresponding estimates for square lattice 
SAWs.

An analysis of the total number of SAWs on the Ammann-Beenker tiling using first
order DAs yields $x_c=0.3647(33)$ and $\gamma=3.14(37)$.
Whereas the critical point estimate is consistent with the previous analysis,
the exponent estimate deviates from the value of $43/32=1.34375$ for fixed origin 
SAWs or the mean number of SAWs.
This phenomenon reflects the fact that the number of Ammann-Beenker patches of 
radius $r$ grows asymptotically as $r^2$. 
(More generally, for aperiodic Delone sets in $\mathbb R^d$ described
by a primitive substitution matrix, the number $N(r)$ of patches of
radius $r$ grows like $N(r)\simeq r^d$ \cite{Le02,LP01}.)
Since the SAW has fractal dimension $4/3$, we expect an asymptotic increase of the
number of SAWs by $n^{2\nu}$, where $\nu=3/4$.
Thus $\gamma=43/32+2\nu=2.84375$.
Data extrapolation is consistent with this value.

The analysis of SAP data follows the same lines.
However, the estimates suffer from large finite size errors due to the low 
number (11) of available coefficients.
First order differential approximants for the mean number of SAPs yield 
$x_c=0.3688(41)$.
We assume $x_c(SAP)=x_c(SAW)$, which has been proven for the square lattice case 
\cite{H61}.

For the critical exponent $\alpha=2+\gamma$, we expect for the mean
number of SAPs by universality that $\alpha=1/2$, being the believed
exact value for the square lattice (and numerically confirmed to very
high precision \cite{J00}).
Due to lack of data it is not possible to give estimates of critical
exponents.  An analysis of the total number of SAPs on the
Ammann-Beenker tiling using first order DAs yields $x_c=0.3587(15)$.

The above analysis has also been applied to the rhombic Penrose tiling data.
We observe qualitatively the same finite size behaviour as for the 
Ammann-Beenker tiling data, though the fluctuations are a bit less pronounced.
In Table \ref{Ta:diffappP} we list estimates for $x_c$ and $\gamma$
obtained by analysing first order differential approximants.  

\begin{table}[ht]
\center{\scalebox{1.0}{\parbox{\textwidth}{
\begin{tabular}{|c|r|r|r|r|r|} \hline
 & mean no. &  $(0,0,0)$ & $(0,0,1)$ & $(0.25,0.5,0)$ & $(0.5,0.5,1)$ \\ \hline
$x_c$ & 0.36322(29) & 0.3621(12)& 0.3613(16)&  0.36248(83)& 0.36347(51) \\ \hline
$\gamma$ & 1.333(26) & 1.28(14)& 1.19(22)& 1.303(83)& 1.387(62)  \\
 \hline
\end{tabular}
}}}
\caption{Estimates of $x_c$ and $\gamma$ for Penrose SAWs.
Numbers in brackets denote the uncertainty in the last two digits.}
\label{Ta:diffappP}
\end{table}

An analysis of the total number of SAWs on the rhombic Penrose tiling using first
order DAs yields $x_c=0.3638(31)$ and $\gamma=2.77(21)$.
The estimate of the critical point is consistent with the estimates from the
other methods of counting.
For the critical exponent, we again expect a value of $\gamma=43/32+2\nu=2.84375$,
which agrees with the extrapolation within numerical accuracy.

For the analysis of SAP data, only 9 series coefficients are available.
First order differential approximants for the mean number of SAPs yield 
$x_c=0.372(11)$.
An analysis of the total number of SAPs on the rhombic Penrose tiling using first
order DAs yields $x_c=0.3590(22)$.
Again, due to lack of data, it is not possible to extrapolate reasonable
estimates for the critical exponent $\alpha$.

\section*{Conclusion}

We extended previous enumerations for self-avoiding walks and polygons 
on the Ammann-Beenker tiling and on the rhombic Penrose tiling and
extracted estimates for the critical point and critical exponent,
using different counting schemes.
It turned out that averaging with respect to the occurrence probability in the whole
tiling leads to the best estimates, whereas data produced by fixing an origin
leads to strong finite size oscillations.
The results support the universality hypothesis that the critical exponents
appear to be the same as for the square lattice, within confidence limits.
For the total number of walks (polygons) we obtain a new exponent reflecting the 
polynomial complexity of the number of patches of the underlying tiling.

Since the results were obtained using the enumeration method of backtracking,
one might ask if more efficient enumeration methods can be applied in order
to substantially increase the length of the series, and hence the
accuracy of the estimates.
Unfortunately, the successful finite-lattice method cannot be applied in this 
case, without substantial development.

On a mathematically rigorous level, it may be possible to show the
equality of the critical points for SAPs and SAWs on quasiperiodic tilings
by appropriately modifying the existing proofs for the hypercubic lattice \cite{H61}.
Furthermore, it would be interesting to carry out an analysis to
determine if random walk behaviour
can be proved for dimensions greater than four \cite{MS93}.

Self-avoiding polygons may also be counted by area.
Since this leads to a three-variable generating function due to the different 
areas of the prototiles, it is tempting to ask whether the scaling behaviour of 
these objects is different from that recently found \cite{RGJ01} for 
self-avoiding polygons on two-dimensional lattices.

\section*{Acknowledgements}

The authors thank Michael Baake for bringing their attention to the above problem
and Daniel Lenz for useful advice on substitution systems. 
CR would like to acknowledge funding by the German Research Council (DFG).
AJG would like to acknowledge funding by the Australian Research Council (ARC).

\end{document}